\documentclass{INTERSPEECH2023}
\usepackage{xcolor}
\usepackage{glossaries}
\usepackage{cite}
\usepackage{multirow}
\usepackage{svg}
\usepackage{subfig}
\usepackage{xspace}
\makeatletter
\DeclareRobustCommand\onedot{\futurelet\@let@token\@onedot}
\def\@onedot{\ifx\@let@token.\else.\null\fi\xspace}
\def\eg{\emph{e.g}\onedot}

\def\etal{\emph{et al}\onedot}
\makeatother

\newacronym{enf}{ENF}{electric network frequency}
\newacronym{snr}{SNR}{signal-to-noise ratio}
\newacronym{rt}{RT}{reverberation time}
\newacronym{dl}{DL}{deep learning}
\newacronym{cnn}{CNN}{convolutional neural network}
\newacronym{crnn}{CRNN}{convolutional recurrent neural network}
\newacronym{rnn}{RNN}{recurrent neural network}
\newacronym{nn}{NN}{neural network}
\newacronym{seq2seq}{seq2seq}{sequence-to-sequence}
\newacronym{rir}{RIR}{room impulse response}
\newacronym{pp}{pp}{percentage points}
\newacronym{fc}{FC}{fully-connected}

\usepackage{booktabs}
\newcommand{\ra}[1]{\renewcommand{\arraystretch}{#1}}

\interspeechcameraready

\title{Point to the Hidden: Exposing Speech Audio Splicing via Signal Pointer Nets}
\name{Denise Moussa$^{1,2}$, Germans Hirsch$^2$, Sebastian Wankerl$^3$, Christian Riess$^2$}
\address{
	$^1$Federal Criminal Police Office (BKA), Germany\\
	$^2$IT Security Infrastructures Lab, Friedrich-Alexander University of Erlangen-Nürnberg, Germany\\
	$^3$Data Science Chair, Julius-Maximilians University of Würzburg, Germany}
\email{\{denise.moussa, christian.riess\}@fau.de}
\begin{document}

\maketitle
 
\begin{abstract}
Verifying the integrity of voice recording evidence for criminal investigations is an integral part of an audio forensic analyst's work. Here, one focus is on detecting deletion or insertion operations, so called audio splicing. While this is a rather easy approach to alter spoken statements, careful editing can yield quite convincing results. For difficult cases or big amounts of data, automated tools can support in detecting potential editing locations. To this end, several analytical and deep learning methods have been proposed by now. Still, few address unconstrained splicing scenarios as expected in practice. With SigPointer, we propose a pointer network framework for continuous input that uncovers splice locations naturally and more efficiently than existing works. Extensive experiments on forensically challenging data like strongly compressed and noisy signals quantify the benefit of the pointer mechanism with performance increases between about $6$ to $10$ percentage points.~\footnote{code: https://www.cs1.tf.fau.de/research/multimedia-security}

\end{abstract}

\section{Introduction}
In today's digital era, more and more speech recordings like voice messages, recorded phone calls or audio tracks of videos are produced and possibly post-processed and shared via the internet. 
Consequently, they often contain important cues for criminal investigations, too. 
With powerful tools, either commercial or free, as for example Audacity~\cite{audacity}, the hurdles for editing operations have become low. 
Forensic audio analysts are thus often assigned to verify the integrity of material relevant to court cases.
Audio splicing (which subsumes deletion, copying and insertion of speech segments) is an effective and easy-to-perform manipulation that violates integrity. 
For example, the sentence \emph{I do not agree} is easily inverted by deleting the signal segment containing \emph{not} and merging the remaining parts. 
Simple post-processing steps, such as saving forgeries using lossy compression, \eg in MP3 format, can further weaken or obscure editing cues. Furthermore, the workload for analysts strongly increases with forgery quality and amount of data. 
Up to now, several methods have been proposed to assist with localising splices in speech material. 
However, they are mostly inapplicable to unconstrained signal characteristics. 
With this work, we address the current limitations and propose a novel and natural approach to audio splicing localisation.

\subsection{Existing Approaches to Audio Splicing Localisation}
\label{sec:related_work}
Audio splicing localisation is mostly targeted with analytical and \gls{dl} based methods that focus on specific features to detect signal inconsistencies. 
By example, some previous works examine the consistency of specific audio formats~\cite{yang2008detecting, cooper2010detecting, xiang2022forensic}, rely on splices of recordings from different devices~\cite{cuccovillo2013audio,baldini2022microphone}, or detect changes of the recording environment's noise levels~\cite{pan2012detecting,yan2021exposing}, acoustic impulse~\cite{capoferri2020speech} or both~\cite{zhao2017audio}.
Others also search for atypical changes in the subtle (and fragile)~\gls{enf}~\cite{esquef2015improved, lin2017supervised, mao2020electric}.
Due to the rise of convincing audio synthesis techniques, several works specifically aim at detecting artificial segments amidst original speech~\cite{zhang2022localizing, zhang2022robust}. 

In practice, forensic analysts are confronted with audio samples from unconstrained sources, which implies, \eg, arbitrary recording parameters, quality, formats, or post-processing operations. So, methods relying on the presence of very specific features might not be applicable if those are not present in certain audio signals.
Indeed, recently, several \gls{dl} approaches with unconstrained feature extraction have been proposed~\cite{jadhav2019audio,patole2021machine,chuchra2022deep, zhang2022aslnet, zengaudio, moussa2022towards}, however many of these methods are still preliminary. Some either target audio splicing detection but omit localisation~\cite{jadhav2019audio, patole2021machine, chuchra2022deep}, another work examines only two fixed splicing patterns, and its generalization remains unclear~\cite{zhang2022aslnet}.
In addition, the  small and non-diverse Free Spoken Digit Dataset~\cite{fsdd} is often used to construct spliced samples~\cite{jadhav2019audio, patole2021machine, chuchra2022deep}, and frequently, material from different speakers instead of one is merged~\cite{jadhav2019audio, patole2021machine, zhang2022aslnet, chuchra2022deep}, which excludes the highly relevant and more difficult-to-detect case of forged statements of one person. Zeng~\etal~\cite{zengaudio} consider more miscellaneous spliced forgeries from one speaker and employ a ResNet-18~\cite{he2016deep} method for chunks of audio spectrogram frames. However, this is only fit for coarse splicing localisation within windows of $32$ to $64$ frames. For frame-level granularity, a \gls{seq2seq} Transformer~\cite{vaswani2017attention} model has been proposed~\cite{moussa2022towards}. It outperforms several \gls{cnn} classifiers on challenging data, but still leaves room for improvement concerning well-made forgeries.
\subsection{Audio Splicing Localisation via Pointer Mechanisms}
\label{sec:pointer_and_s2s}
We propose a major improvement over existing methods for audio splicing localisation by regarding this task as a pure pointing problem. Pointers predict a conditional probability distribution over elements of a sequence and were originally designed for approximating combinatorial optimization problems~\cite{vinyals2015pointer}. They can thus directly locate parts of the input series, in contrast to traditional \gls{seq2seq} networks,  where a mapping to a fixed set of target tokens is learned. Pointer mechanisms were recently also integrated into the Transformer~\cite{vaswani2017attention} architecture, where mixtures of pointer and token generation components solve natural language processing tasks~\cite{enarvi2020generating, yan2021unified}.
In our case, we want the \gls{nn} to indicate splice locations by pointing to the respective input signal positions. This appears more natural and efficient than step-wise classifying segments of fixed size~\cite{zengaudio} or learning a mapping to a fixed vocabulary~\cite{moussa2022towards}. Existing pointer methods however operate on categorical input. We thus design SigPointer, a Transformer~\cite{vaswani2017attention} based pointer network for continuous input signals. We benchmark against existing works on audio splicing localisation and analyse the influence of our network's components. SigPointer proves to perform best on forensically challenging data, both under seen and unseen quality degradations.

\section{Proposed Method}
In this section, we describe our proposed pointing method for signals, as well as training strategies and datasets.

\subsection{SigPointer for Continuous Input Signals}
\label{subsec:sigpointer}
\begin{figure}
	\scalebox{0.6}{	\includesvg{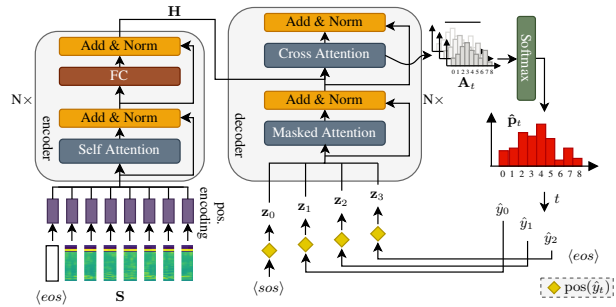}}
	\caption{SigPointer model for locating splices in audio signals}\label{fig:model}
\end{figure}

We define audio splicing localisation as a pointing task. The input to our encoder-decoder network (Fig.~\ref{fig:model}) is a time series of signal representation vectors with length $N$, formally $\mathbf{S} = [\mathbf{s}_0, \mathbf{s}_1, \cdots , \mathbf{s}_N]$. Additionally, $\mathbf{s}_{-1}$ denotes $\langle eos \rangle$, a special token to which the pointer mechanism can point to denote that decoding is finished~\cite{yan2021unified}. It is omitted in the following definitions for simplicity. The encoder part of our network (left) mostly aligns with the original Transformer~\cite{vaswani2017attention} model. It is made of a stack of layers, each implementing multi-head self-attention followed by a \gls{fc} layer and layer-normalisation. However, unlike existing pointer methods on categorical data (cf. Sec.~\ref{sec:pointer_and_s2s}), we skip learning input embeddings and feed the raw data into the network, since we already operate on continuous, dense data. The encoded representation $\mathbf{H} = [\mathbf{h}_0, \mathbf{h}_1, \cdots, \mathbf{h}_N]$ of $\mathbf{S}$, where $\mathbf{h}_n \in \mathbb{R}^l$ with latent size $l$, then serves as input to the decoder (middle). The decoder solely consists of a stack of  multi-head attention layers~\cite{vaswani2017attention}. Per time step $t$, it takes $\mathbf{H}$ and one position vector $\mathbf{z}_{t^{*}} \in \mathbb{R}^l$ per previously predicted $\hat{y}_{t^{*}}  \in \mathcal{I} =  \{0, 1, \cdots, N\}$ with $t^* < t$ to indicate the next splice location $\hat{y}_t$. More details about this decoding process are given in the following two paragraphs. 

 \textbf{Pointer Mechanism} To predict the full output index  sequence  $\mathbf{\hat{y}}$, we need the conditional probability distribution over all input positions $P(\mathbf{\hat{y}}|\mathbf{S}) = \prod_{t=0}^{T} P(\hat{y}_t | \mathbf{H}, \mathbf{Z})$. Thus, for each time step, we extract the cross-attention scores  between $\mathbf{H}$ and $\mathbf{Z}$ from the last decoder layer for all $M$ attention heads, yielding $\mathbf{A}^{M\times|\mathbf{S}|}_t$, with $a_{m,s} \in \mathbb{R}$ (Fig.~\ref{fig:model}, right). Unlike related Transformer pointer methods, we do not mix pointer and token generation tasks~\cite{enarvi2020generating, yan2021unified}, so all the model's attention heads can be reserved for computing one final pointer result. We thus average and normalize all attention heads' values to yield the discrete index probability distribution as $\mathbf{\hat{p}_t} = \operatorname{softmax}( {\left.\overline{\mathbf{A}_t}\right|_1^M})$. During inference, $\hat{y}_t = \operatorname{argmax}(\mathbf{\hat{p}_t})$ is computed to yield the final splice point or $\langle eos \rangle$ if no (further) splice point is detected.
 
\textbf{Slim Decoding} Existing \gls{seq2seq} (pointer) approaches on categorical data traditionally use the semantic of the previously decoded elements $\hat{y}_{t^*}$ to decode the next $\hat{y}_{t}$ from $\mathbf{H}$. Hereby, $\hat{y}_{t^*}$ is projected to latent size $l$ with an embedding layer $E_m$ and positional encodings are added to yield the final representation. Contrary, we only use sinusoidal encoding vectors~\cite{vaswani2017attention} $\mathbf{z}_{t^{*}}$ =  $\mathbf{e}_{t^*}$ for each $\hat{y}_{t^*}$.
The vectors $\mathbf{z}_{t^{*}}$ thus only preserve the relative ordering and number of already decoded output items. In several tests, we observed that this slim information is sufficient, since training some $E_m$ and reusing $y_{t^*}$ as decoder target input did not show any advantage. In fact, the performance even slightly degraded. We reason that this is due to sparser available context between sequence elements compared to approaches on discrete data, \eg, word tokens from a natural language~\cite{enarvi2020generating, yan2021unified}. Differently from the latter, splicing points carry few syntactical information (locations) and are generally not expected to have rich dependencies among each other. So, previously found splice locations do not provide clues about future ones.

\subsection{Choosing Model and Training Parameters}
\label{sec:hyperparams}
We conduct a parameter search of $200$ trials with optuna~\cite{optuna_2019} (v.2.10) and PyTorch~\cite{PaszkePyTorch2019} (v.1.10.2) on $4$ consumer GPUs. The search space encompasses $n_e, n_d \in [1, 8]$ encoder and decoder layers, $h \in \{1,3,9\}$ heads (divisors of latent size $l=279$), $f_d \in [2^7, 2^{11}]$ for the \gls{fc} layer dimension, dropout $d \in [0.01, 0.05, \cdots, 0.3]$ within the attention layers, learning rate $l_r \in [1e^{-5}, 5e^{-5}, \cdots 5e^{-3}]$ for the Adam optimizer and batch size $b \in [32, 64, 128, 256, 350]$. We choose the best configuration $\mathcal{C}: (n_e, n_d, h, f_d, d, l_r, b) = (7, 1, 9, 2^7, 0.1, 5e^{-4} , 350)$ for our network and use Glorot weight initialisation~\cite{glorot2010understanding}.
SigPointer converges significantly better with regression as opposed to classification losses.
 The cosine distance loss  $d_c = \left. \overline {\frac{\mathbf{\hat{p}}_t \cdot \mathbf{p}_t }{ \lVert \mathbf{\hat{p}}_t \rVert  \lVert \mathbf{p}_t \rVert }}\right|_1^T $ proves to be most stable, where $\mathbf{p}_t$ are the one-hot encoded target splice indices. Note that a more exact metric for our task is the mean absolute error between the predicted positions $\mathbf{\hat{y}}$ and targets $\mathbf{y}$. Since it guided training better but is non-differential, we only use it as validation loss.

\subsection{Datasets and Training Strategy}
\label{sec:training_strategy}
All datasets are generated using our previously introduced pipeline~\cite{moussa2022towards} and cover forensically challenging scenarios. We use the anechoic ACE dataset~\cite{eaton2016estimation} as source set, such that the model does not adapt to  unintended, acoustic side channels in forged samples. Samples between $3$\,s and $45$\,s are created from $n_s$ source segments of the same speaker from the same or different simulated environments. Post-processing that may obscure splicing is applied in the form of additive Gaussian noise and either single AMR-NB or MP3 compression, all in randomly sampled strength. Adding noise to a signal can serve as an easy method to mask tampering, just as compression which however may also be unintentionally introduced by re-saving the edited version or sharing via social media services. The final feature representation with a time resolution of $500$\,ms is a concatenation of the Mel spectrogram, MFCCs and spectral centroids, yielding the feature dimension $l = 279$. For more details about the generation pipeline we refer to the respective work~\cite{moussa2022towards}.

For training, we employ curriculum learning~\cite{bengio2009curriculum} in a three-stage process with a train/validation split of 500k/30k for each stage. Training is conducted upon model convergence with $100$ epochs and a patience of $20$ epochs. The first dataset covers samples with $n\in [0, 1]$ splices and no post-processing, the second extends to post-processing and the third includes both multi-splicing with $n \in [0, 5]$ and post-processing.
As in existing work~\cite{moussa2022towards}, we employ cross-dataset testing and generate our test sets from the Hi-Fi TTS set~\cite{bakhturina21_interspeech} as described in Sec.~\ref{sec:experiments}.

\begin{table*}[t]\centering
	\ra{0.9}
	\caption{Model size and performance (mean $\pm$ SD of $5$ training runs) on the Hi-Fi TTS test set with $n \in [0,5]$ splices per sample. Jaccard Index $J$ and recall $R$ are evaluated for exact localisation ($\operatorname{Bin}=1$) and w.r.t $f\in[2,3,4]$ frame binning of the input signal. }\label{tab:tts-05}
	\resizebox{\textwidth}{!}{%
	\begin{tabular}{@{}l@{\hskip 0.05cm}r@{\hskip 0.1cm}l@{\hskip 0.1cm} l@{\hskip 0.1cm}l@{\hskip 0.1cm}l@{\hskip 0.2cm} l@{\hskip 0.1cm}l@{\hskip 0.1cm}l@{\hskip 0.2cm} l@{\hskip 0.1cm}l@{\hskip 0.1cm}l@{\hskip 0.2cm} l@{\hskip 0.1cm}l@{\hskip 0.1cm} @{}}\toprule[1pt]
		
		\multirow{2}{*}{Model}&	\multicolumn{1}{l}{\multirow{2}{*}{Params}}&\phantom{l} & \multicolumn{2}{c}{$\operatorname{Bin}=1$}   &&                  \multicolumn{2}{c}{$\operatorname{Bin}=2$}&                      & \multicolumn{2}{c}{$\operatorname{Bin}=3$} &                   & \multicolumn{2}{c}{$\operatorname{Bin}=4$}                     \\  \cmidrule{4-5} \cmidrule{7-8} \cmidrule{10-11}  \cmidrule{13-14}
		&&& \multicolumn{1}{c}{$J$} & \multicolumn{1}{c}{$R$} && \multicolumn{1}{c}{$J$} & \multicolumn{1}{c}{$R$} && \multicolumn{1}{c}{$J$} & \multicolumn{1}{c}{$R$} && \multicolumn{1}{c}{$J$} & \multicolumn{1}{c}{$R$} \\ \midrule[1pt] 
		Jadhav~\cite{jadhav2019audio}                &205.15 M && 0.2189$\scriptstyle\pm0.004$                & 0.2369$\scriptstyle\pm0.004$                && 0.2547$\scriptstyle\pm0.004$                & 0.2768$\scriptstyle\pm0.004$                && 0.2745$\scriptstyle\pm0.004$                & 0.2982$\scriptstyle\pm0.004$                && 0.2952$\scriptstyle\pm0.004$                & 0.3191$\scriptstyle\pm0.004$                \\
		Chuchra~\cite{chuchra2022deep}               &134.77 K&& 0.2964$\scriptstyle\pm0.004$                & 0.3089$\scriptstyle\pm0.004$                && 0.3280$\scriptstyle\pm0.004$                & 0.3373$\scriptstyle\pm0.005$                && 0.3445$\scriptstyle\pm0.005$                & 0.3509$\scriptstyle\pm0.005$                && 0.3533$\scriptstyle\pm0.005$                & 0.3594$\scriptstyle\pm0.005$                \\ \vspace{0.15cm}
		Zeng~\cite{zengaudio}                  &11.17 M&& 0.3274$\scriptstyle\pm0.003$                & 0.3867$\scriptstyle\pm0.002$                && 0.4249$\scriptstyle\pm0.003$                & 0.4921$\scriptstyle\pm0.004$                && 0.4734$\scriptstyle\pm0.004$                & 0.5413$\scriptstyle\pm0.005$               && 0.5162$\scriptstyle\pm0.004$                & 0.5831$\scriptstyle\pm0.005$                \\ 
		Transf. enc.        &17.51 M&& 0.4377$\scriptstyle\pm0.005$                 & 0.4855$\scriptstyle\pm0.007$                 && 0.5386$\scriptstyle\pm0.005$                 & 0.5825$\scriptstyle\pm0.008$                 && 0.5839$\scriptstyle\pm0.006$                 & 0.6231$\scriptstyle\pm0.009$                 && 0.6231$\scriptstyle\pm0.006$                 & 0.6574$\scriptstyle\pm0.010$                 \\ 
		Moussa~\cite{moussa2022towards}                 &7.62 M&& 0.4198$\scriptstyle\pm0.005$                 & 0.4747$\scriptstyle\pm0.008$                 && 0.5263$\scriptstyle\pm0.008$                 & 0.5794$\scriptstyle\pm0.011$                 && 0.5679$\scriptstyle\pm0.010$                 & 0.6187$\scriptstyle\pm0.014$                 && 0.6123$\scriptstyle\pm0.011$                 & 0.6595$\scriptstyle\pm0.015$                 \\
		SigPointer\textsubscript{$\mathcal{C}_M$}              &7.57 M&& \emph{0.4670}$\scriptstyle\pm0.003$                & \emph{0.5202}$\scriptstyle\pm0.003$                && \emph{0.5767}$\scriptstyle\pm0.003$                & \emph{0.6265}$\scriptstyle\pm0.006$                && \emph{0.6160}$\scriptstyle\pm0.003$                & \emph{0.6620}$\scriptstyle\pm0.006$                && \emph{0.6648}$\scriptstyle\pm0.004$                & \emph{0.7061}$\scriptstyle\pm0.008$       \\
		SigPointer*     &3.40 M&& \textbf{0.5184}$\scriptstyle\pm0.006$       & \textbf{0.5719}$\scriptstyle\pm0.011$       && \textbf{0.6228}$\scriptstyle\pm0.012$       & \textbf{0.6675}$\scriptstyle\pm0.018$       && \textbf{0.6607}$\scriptstyle\pm0.016$       & \textbf{0.7002}$\scriptstyle\pm0.022$      && \textbf{0.6977}$\scriptstyle\pm0.019$       & \textbf{0.7322}$\scriptstyle\pm0.025$\\
		\bottomrule[1pt]   
	        
	\end{tabular}
}
	
\end{table*}
\section{Experiments}
\label{sec:experiments}
We evaluate our proposed method on forensically challenging data and test against four existing methods and two custom baselines to analyse the benefits of the pointer framework.

\subsection{Baseline Models}
\label{sec:baselines}
We train each \gls{nn} as described in Sec.~\ref{sec:training_strategy}, where all but the pointer approaches employ the BCE instead of the cosine loss.

\textbf{CNNs} Most existing works rely on \glspl{cnn} and natively have too strong limitations to be directly used for our task (cf. Sec.~\ref{sec:related_work}).  We reimplement three approaches~\cite{jadhav2019audio, chuchra2022deep, zengaudio}. The first two frameworks~\cite{jadhav2019audio, chuchra2022deep} only cover binary classification of spliced vs. non-spliced signals with custom \gls{cnn} models, while Zeng~\etal~\cite{zengaudio} propose localisation with a ResNet-18~\cite{he2016deep} classifier.
They employ a sliding window approach over chunks of frames, where the window stride $s=1$ accounts for exact (frame-level) detection. The classifier's decision per frame is inferred from averaged probabilities of multiple windows. This approximative method is however only fit for larger $s$~\cite{zengaudio}. Our reimplementation confirms long training times and poor detection on frame level, so we use a custom splice localisation strategy for \glspl{cnn}. It is based on  our previously introduced framework~\cite{moussa2022towards}, where baseline \glspl{cnn} are extended to classify at maximum $n$ splicing positions with $n_o = n$ output layers. This however shows poor performance for $ n > 1$. Instead, we set $n_o$ to the maximum expected number of input frames, $90$ in our case, and perform a binary classification per frame which proves to be more stable w.r.t. higher $n$. Splitting in smaller segments accounts for signals with lengths exceeding $n_o$.
We exclude two methods because of no details on the model specifications~\cite{patole2021machine} and very restrictive splicing assumptions where a relaxation to our more unconstrained task is not straight forward~\cite{zhang2022aslnet}. 

\textbf{Seq2Seq Transformer} We also re-train our \gls{seq2seq} model from previous work on multi-splicing localisation~\cite{moussa2022towards}. Given a signal, the encoder outputs its latent representation from which a sequence of splicing points is decoded step-by-step using a fixed vocabulary set.
   
\textbf{SigPointer\textsubscript{$\mathcal{C}_M$}} For a direct comparison of the pointer against the \gls{seq2seq} framework, we instantiate SigPointer with the Transformer configuration $\mathcal{C}_M$ in~\cite{moussa2022towards}.

\textbf{Transformer encoder} SigPointer employs autoregressive decoding (cf. Sec.~\ref{subsec:sigpointer}). To quantify its influence we test against the capacity of a plain, non-autoregressive Transformer encoder. 
Thereby, we project the encoder memory $\mathbf{H}^{l \times N}$ (Fig.~\ref{fig:model}) to size $2 \times N$ to perform a per frame classification as for the \gls{cnn} baselines. For a fair comparison we again conduct a hyper-parameter search (Sec.~\ref{sec:hyperparams}), but double the search space for the number of encoder layers to $n_e \in [1,16]$ to account for the missing decoder capacity. The best model configuration yields $\mathcal{C}: (n_e, h, f_d, d, l_r, b) = (12, 9, 2^{11}, 0.1, 1e^{-4} , 64)$.

\subsection{Performance Metrics}
We report the average Jaccard index $J$ expressing the similarity of prediction $\mathbf{\hat{y}}$ and ground truth $\mathbf{y}$ as intersection over union \mbox{$J =\frac{|\mathbf{\hat{y}} \cap \mathbf{y}|}{|\mathbf{\hat{y}} \cup \mathbf{y}|}$}, as well as average recall $R =  \frac{|\mathbf{\hat{y}} \cap \mathbf{y}|}{|\mathbf{y}|}$ of $5$ training runs with different seeds. Note that the order of the predicted points is irrelevant for both metrics. We evaluate both exact localisation and coarser granularity by binning the input signal by $f \in [1,2,3,4]$ frames, denoted as $\operatorname{Bin}=f$.

\subsection{Evaluation of the Pointer Mechanism}
\label{subsec:eval_pointer}
For this experiment, we generate a dataset of $30$k samples with uniformly sampled  $n \in [0, 5]$ splicing positions from the Hi-Fi TTS test pool~\cite{bakhturina21_interspeech}, including single compression and noise post-processing as described in Sec.~\ref{sec:training_strategy}. The size and performance results of all models are listed in Tab.~\ref{tab:tts-05}.
 The \gls{cnn} methods (rows $1$ to $3$) are inferior to the Transformer-based approaches (rows $4$ to $7$). Jadhav~\etal's~\cite{jadhav2019audio} large but very shallow network performs worst, followed by Chuchra~\etal's~\cite{chuchra2022deep} small but deeper model with $12$ layers and the best and deepest ResNet-18 \gls{cnn} baseline.
Solving the same  classification task (Sec.~\ref{sec:baselines}) with the Transformer encoder greatly improves splicing localisation. It also slightly surpasses the best proposed \gls{seq2seq} model from related work~\cite{moussa2022towards}. However, it also uses about $2.3$ as much trainable parameters. Training the \gls{seq2seq} model in our pointer framework (SigPointer\textsubscript{$\mathcal{C}_M$}) demonstrates the benefit of our proposed approach.
The missing \gls{seq2seq} vocabulary mapping component slightly reduces the network size, still the Jaccard index and recall increase by approximately $5.0$ \gls{pp} and $4.8$ \gls{pp} and outperform both the Transformer encoder and the original \gls{seq2seq} model. 
We achieve the best performance with optimized hyper-parameters (cf. Sec.~\ref{sec:hyperparams}), yielding the even smaller SigPointer*. Notably, the decoder is reduced to $1<5$ layers compared to SigPointer\textsubscript{$\mathcal{C}_M$} which suffices for our slim decoding strategy (cf. Sec~\ref{subsec:sigpointer}). The advantage of about $5.1$ \gls{pp} and $5.2$ \gls{pp} to  SigPointer\textsubscript{$\mathcal{C}_M$} with $J = 0.5184 > 0.4670$ and $R = 0.5719 > 0.5202$ for $\operatorname{Bin}=1$ steadily decreases, reaching $3.8$~\gls{pp} ($J$) and $3.0$~\gls{pp} ($R$) for $\operatorname{Bin}=4$. We thus assume that SigPointer\textsubscript{$\mathcal{C}_M$} is only slightly less sensitive to splicing points than the optimized SigPointer* but notably less exact in localisation. 

\subsection{Influence of Splices per Input}
\label{subsec:eval_no_splices}

\begin{figure*}[t]
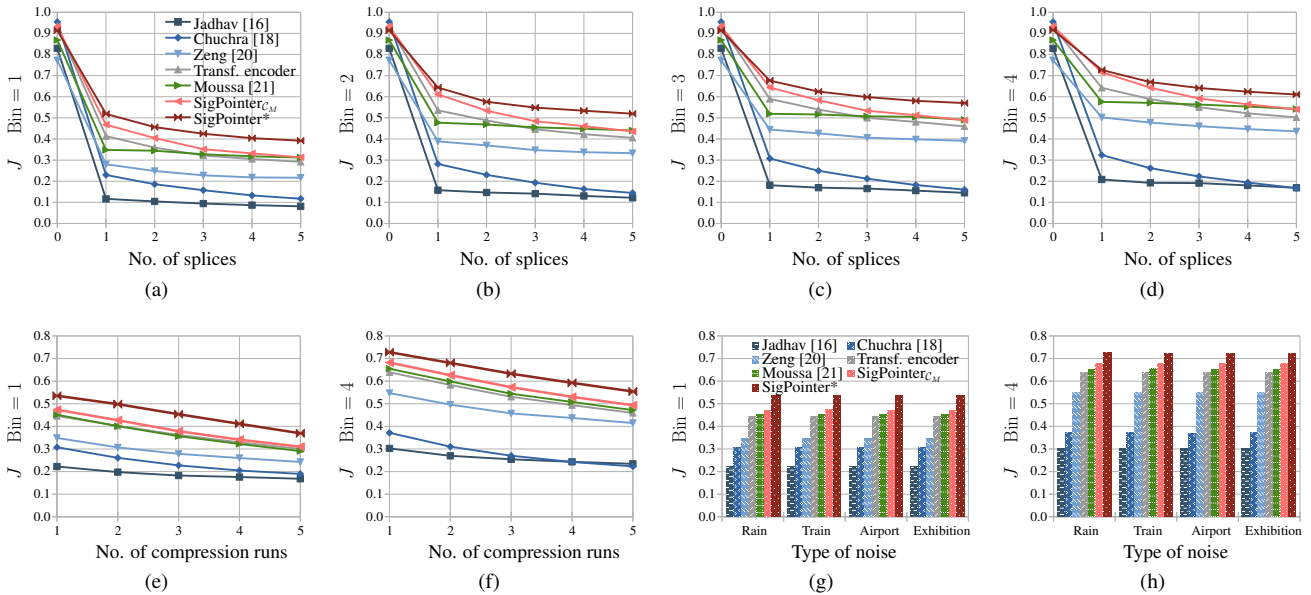

	\newcommand{\thspace}{0.40cm}
	\newcommand{\imgscale}{0.42}
	\centering
	\subfloat[\label{subfig:tts-per-splice-a}]{\scalebox{\imgscale}{	\includesvg{images/tts_05_bin1_new.svg}}}\hspace{\thspace}
	\subfloat[\label{subfig:tts-per-splice-b}]{\scalebox{\imgscale}{	\includesvg{images/tts_05_bin2_new.svg}}}\hspace{\thspace}
	\subfloat[\label{subfig:tts-per-splice-c}]{\scalebox{\imgscale}{	\includesvg{images/tts_05_bin3_new.svg}}}\hspace{\thspace}
	\subfloat[\label{subfig:tts-per-splice-d}]{\scalebox{\imgscale}{	\includesvg{images/tts_05_bin4_new.svg}}}\hspace{\thspace}\\
		\subfloat[\label{subfig:compr_bin1}]{\scalebox{\imgscale}{	\includesvg{images/compr_bin1_new.svg}}}\hspace{\thspace}
	\subfloat[\label{subfig:compr_bin4}]{\scalebox{\imgscale}{	\includesvg{images/compr_bin4_new.svg}}}\hspace{\thspace}
	\subfloat[\label{subfig:noise_bin1}]{\scalebox{\imgscale}{	\includesvg{images/noise_bin1_new.svg}}}\hspace{\thspace}
	\subfloat[\label{subfig:noise_bin4}]{\scalebox{\imgscale}{	\includesvg{images/noise_bin4_new.svg}}}
	\caption{Jaccard index J (mean of 5 training runs) for  $n \in [0,5]$ splices (Figure~\ref{subfig:tts-per-splice-a}-\ref{subfig:tts-per-splice-d}) and robustness towards out-of-distribution multi-compression and real noise post-processing (Fig.~\ref{subfig:compr_bin1}-\ref{subfig:noise_bin4}). The pointer framework (red tones) is clearly superior in all tests.}
	\label{fig:tts-per-splice}
\end{figure*}

In Fig.~\ref{fig:tts-per-splice} we report the Jaccard coefficients on our Hi-Fi TTS test set w.r.t.\ the number of splice positions per sample.
Evidently, all models recognise non-spliced inputs relatively well, while localising actual splice positions correctly proves to be more difficult on this challenging dataset (Fig.~\ref{subfig:tts-per-splice-a}-\ref{subfig:tts-per-splice-d}). The best performing SigPointer* (red) achieves $J = 0.5187$ for single splices and still $J = 0.3918$ for $n=5$ splices. When allowing coarser signal binning (Fig.~\ref{subfig:tts-per-splice-b}-\ref{subfig:tts-per-splice-d}), the performance increases considerably up to $J = 0.7268$ and $J = 0.6104$ for $n = 1,5$ and $\operatorname{Bin}=4$ (Fig.~\ref{subfig:tts-per-splice-d}). As stated in Sec.~\ref{subsec:eval_pointer}, the less accurate SigPointer\textsubscript{$\mathcal{C}_M$} (orange) benefits from coarser bins, but cannot outperform SigPointer*. The Transformer encoder (grey) performs well for $n=1,2$, but drops below the Transformer \gls{seq2seq} model's~\cite{moussa2022towards} (green) performance for $n\geq3$. 
The \glspl{cnn} (blue) are weakest, where especially the custom models~\cite{jadhav2019audio, chuchra2022deep} exhibit low sensitivity to splicing and thus cannot profit from coarser signal binning (Fig.~\ref{subfig:tts-per-splice-b}-\ref{subfig:tts-per-splice-d}). In summary, SigPointer* outperforms all baselines in terms of sensitivity to splicing and exactness of localisation, despite its small model size of $3.4$\,M parameters.

\subsection{Robustness to Complex Processing Chains}
We test the generalization ability of the models trained only on single compression and additive Gaussian noise post-processing to even stronger obscured splicing points. We thus run the robustness experiments from our previous work on the available test sets \`a 10k samples from Hi-Fi TTS~\cite{moussa2022towards}. They consist of $0$ to $5$ times spliced samples subjected to $n_c \in [1,5]$ AMR-NB or MP3 compression runs ($5$ test sets) and additive real noise post-processing ($4$ test sets). Both compression and noise strength are randomly sampled~\cite{moussa2022towards}.
Figures~\ref{subfig:compr_bin1}-\ref{subfig:noise_bin4} show the results for $\operatorname{Bin}=1,4$.
 Most previously described trends in model performance (Sec.~\ref{subsec:eval_pointer} and Sec.~\ref{subsec:eval_no_splices}) also show in this robustness experiment. However, the  \gls{seq2seq} Transformer~\cite{moussa2022towards} this time surpasses the simpler Transformer encoder and shows slightly better robustness in all experiments. SigPointer* again outperforms all models. Compared to the best existing model~\cite{moussa2022towards} (green), for $\operatorname{Bin}=1$ it increases localisation ability by (on average) $8.4$ \gls{pp} for multi compression and $9.1$ \gls{pp} for additive real noise (Fig.~\ref{subfig:compr_bin1}, Fig.~\ref{subfig:noise_bin1}). Surprisingly, despite  differing complexity, the type of noise has little influence on the performance.

\subsection{Limitations}
In our tests, SigPointer models are more sensitive to weight initialisation compared to the other considered \glspl{nn}. 
The number of epochs until the model converges can thus vary strongly, so early pruning of weak runs is recommended in practice.

 Also note that one design advantage of our model can be a pitfall in practice. SigPointer can process signals of arbitrary length, contrary to classifiers that are constrained by their number of output layers or \gls{seq2seq} methods that are indirectly limited by their learned output vocabulary mapping (cf. Sec.~\ref{sec:baselines}). However, we empirically observed that the pointer adapts to signal lengths seen in training and does barely search for splices out of known ranges. The behaviour of inherent adaptation to problem sizes is already known from literature~\cite{kool2018attention,anil2022exploring}. To mitigate this issue, we thus strongly recommend cutting test samples into multiple separate segments that fit into the training distribution, as 
it was also done for the comparison methods.

\section{Conclusions}
With SigPointer, we present a novel and more natural approach to the task of audio splicing localisation with the help of pointer mechanisms. Our focus is on aiding with difficult-to-detect splice positions that pose by example a problem in forensic analysts' daily work. In several tests on in- and out-of-distribution data, we quantified the advantage of our pointer framework for continuous signals and outperform existing approaches by a large margin, even with a much smaller model size.

\bibliographystyle{IEEEtran}
\bibliography{mybib}

\end{document}